\documentclass[twocolumn]{aastex63}

\newcommand{\hxmt}{{Insight-HXMT\/}}

\def\swift{{Swift}}

\def\aql{Aql X--1}
\def\nicer{{NICER}}
\received{2021 June 27}
\revised{2021 August 16}
\accepted{2021 August 17}
\shorttitle{Aql X-1}
\shortauthors{Li et al.}
\begin{document}

\title{Discovery of transition from marginally stable burning to unstable burning after a superburst in Aql X-1}

\correspondingauthor{Zhaosheng Li}
\affiliation{Key Laboratory of Stars and Interstellar Medium, Xiangtan University, Xiangtan 411105,
Hunan, China}
\email{lizhaosheng@xtu.edu.cn}

\correspondingauthor{Yuanyue Pan}
\affiliation{Key Laboratory of Stars and Interstellar Medium, Xiangtan University, Xiangtan 411105,
Hunan, China}
\email{panyy@xtu.edu.cn}

\author{Zhaosheng Li}
\affiliation{Key Laboratory of Stars and Interstellar Medium, Xiangtan University, Xiangtan 411105,
Hunan, China}

\author{Yuanyue Pan}
\affiliation{Key Laboratory of Stars and Interstellar Medium, Xiangtan University, Xiangtan 411105,
Hunan, China}

\author{Maurizio Falanga}
\affiliation{International Space Science Institute (ISSI), Hallerstrasse 6, 3012 Bern,
Switzerland}

%% Note that the \and command from previous versions of AASTeX is now
%% depreciated in this version as it is no longer necessary. AASTeX 
%% automatically takes care of all commas and "and"s between authors names.

%% AASTeX 6.3 has the new \collaboration and \nocollaboration commands to
%% provide the collaboration status of a group of authors. These commands 
%% can be used either before or after the list of corresponding authors. The
%% argument for \collaboration is the collaboration identifier. Authors are
%% encouraged to surround collaboration identifiers with ()s. The 
%% \nocollaboration command takes no argument and exists to indicate that
%% the nearby authors are not part of surrounding collaborations.

%% Mark off the abstract in the ``abstract'' environment. 
\begin{abstract}

Superbursts are long duration, rare, and extremely energetic thermonuclear explosions of neutron star low-mass X-ray binaries (NS LMXBs), which are proposed to be due to unstable carbon ignition. We report the superburst properties and its consequences from \aql\ observed by Neutron Star Interior Composition Explorer (\nicer), the Monitor of All-sky X-ray Image, \swift\ and  Insight Hard X-ray Modulation Telescope (\hxmt) on the Modified Julian Day 59130.7. We find two, faint type I X-ray bursts 9.44 days after the superburst with a short recurrence time of 7.6 minutes, which is the most accurate measurement of the quenching time in all NS LMXBs with observed superbursts. We also discovered mHz quasi-periodic oscillations in the frequency range 2.7--11.3 mHz immediately after the superburst, before and after the resumption of the first type I X-ray burst from \nicer, \swift\ and \hxmt\ observations. For the first time, we observed the transition from superburst, via marginally stable burning to unstable burning in NS LMXBs. We compared the quenching time and the recurrence time of type I X-ray bursts with simulations.   

\end{abstract}

%% Keywords should appear after the \end{abstract} command. 
%% See the online documentation for the full list of available subject
%% keywords and the rules for their use.

%% From the front matter, we move on to the body of the paper.
%% Sections are demarcated by \section and \subsection, respectively.
%% Observe the use of the LaTeX \label
%% command after the \subsection to give a symbolic KEY to the
%% subsection for cross-referencing in a \ref command.
%% You can use LaTeX's \ref and \label commands to keep track of
%% cross-references to sections, equations, tables, and figures.
%% That way, if you change the order of any elements, LaTeX will
%% automatically renumber them.
%%
%% We recommend that authors also use the natbib \citep
%% and \citet commands to identify citations.  The citations are
%% tied to the reference list via symbolic KEYs. The KEY corresponds
%% to the KEY in the \bibitem in the reference list below. 

\section{Introduction} \label{sec:intro}

Neutron star low-mass X-ray binary (NS LMXB) is  composed of a NS accreting matter from its Roche-lobe filling low mass companion star.
%Type I X-ray burst can be triggered on the NS surface due to unstable thermonuclear burning of the hydrogen and helium.
The accreted hydrogen and/or helium can be depleted via stable, unstable, or marginally stable burning processes, which depend on the accretion rate, surface temperature, and abundance \citep[see][for reviews]{Strohmayer06,Galloway21}. The unstable thermonuclear burning of the hydrogen and helium, also known as type I X-ray burst, usually has the duration of $\sim 10-100$ s with a typical energy release of $10^{39}$ erg, recurs from few hours to days, and ignites at a column depth of $y\approx10^8~ \mathrm{g~cm^{-2}}$ \citep[see][]{Lewin93,Galloway08,Galloway20}. There were a few instances of NS LMXBs that exhibited intermediate duration bursts, which can emit $\sim 10^{41}~{\rm erg}$ in $\sim10^2-10^3$ s \citep[e.g.,][]{Falanga08}. In more rare cases, superbursts have been identified from the total energy $\sim 10^{42}$ erg and the duration of $>10^3$ s, which are believed due to burning carbon at an ignition column depth $\sim10^3$ times deeper than typical type I X-ray bursts \citep{Cumming01,Strohmayer02}. The recurrence time of superbursts ranges from  days to years \citep[e.g.,][]{Serino16,Zand17}. 

Between stable and unstable burning, marginally stable burning on the NS surface manifests as the mHz of quasi-periodic oscillations (QPOs) of the X-ray light curves. Several atoll sources, including 4U 1608--52, 4U 1636--53, and \aql, have been discovered as mHz of QPOs in the frequency range $\sim4$--15 mHz \citep{Revnivtsev01, Mancuso21}, corresponding to the timescales of 1--4 minutes.  \citet{Heger07} explains the mHz of QPOs in multizone numerical models
of the NS envelope and proposed that the oscillation frequency depends on the accreted hydrogen fraction and the NS surface gravity, suggesting another way to probe the still unknown NS equation of state.

% The decay of a superburst is well modelled with two power-law component. The first stage, $L\sim t^{-}$.
Superburst provides a unique way to understand the transition of stable, via possible marginally stable, to unstable burning of accreted matter on the NS surface. \citet{Keek11} and \citet{Keek12} simulated the generation of superbursts, and discussed the consequences of NS surface cooling. After a superburst, the NS envelope is sufficiently hot, and all accreted hydrogen and helium are burning stably. Afterwards, the envelope cools down, and the marginally stable burning may occur. As the envelope continues to cool, weak type I X-ray bursts with short recurrence times return. \citet{Cumming04a} estimated the quenching time of one to several weeks after a superburst. However, for NS accreting matter with a solar composition, \citet{Keek12} proposed that the quenching time is five times shorter compared with those from \citet{Cumming04a}, if both $3\alpha$ and the CNO breakout reactions contribute to the stability of thermonuclear burning.  Due to the lack of contiguous observations and the rare occurrence of superbursts, the transition of burning behavior has not been observed, and only upper limits of the quenching time have been constrained \citep[see e.g.,][]{Keek12}. 

% Type I X-ray burst can be used to probe the interactions between the burst emission and its surrounding accretion disk and corona. 
% Careful analysis of the burst spectrum shows that the type I X-ray burst spectra are well fitted by a blackbody with a temperature $T_{\rm bb}\sim 0.5-3 $ keV. However, several distortions of the burst spectrum have been observed. The persistent emission can be enhanced during a burst due to the Poynting-Robertson effect \citep{Zand13}. The reflection emission from the photoionized accretion disk has been observed from the superbursts in 4U 1820--30 \citep{Ballantyne04}, 4U 1636--536, and IGR J17062--6143 \citep{Keek14, Keek17}. Photoionization absorption features have also been identified in 4U 0614+091, 4U 1722--30, and 4U1820--30 \citep{Zand10},  HETE J1900.1--2455 \citep{Kajava17} and GRS 1747--312 \citep{Li18} due to the bound--bound or bound--free transitions of the heavy elements produced in the burning ashes \citep{Weinberg06}.

The transient NS LXMB  \aql\ is composed of a fast spinning pulsar of $\sim$550.27 Hz and a main-sequence K4 spectral-type companion \citep{Callanan99,Chevalier99,Mata17,Casella08}. \aql\ exhibited frequent outbursts and type I X-ray bursts \citep{Campana98,Campana13,Galloway08,Li17,Keek18}. Two superbursts from \aql\ have been observed by the Monitor of All-sky X-ray Image (MAXI) on the Modified Julian Day (MJD) 56493.3 (2013 July 20) and 59130.7 (2020 October 8), which indicates the recurrence time of $\approx7.2$ yr \citep{Serino16, Iwakiri20}. The second superburst has been partially observed by  Neutron Star Interior Composition Explorer (\nicer), which is also its first superburst observation.     

In this work, we will report the \nicer\ and MAXI results of the second superburst from \aql\ during its 2020 outburst. We introduce the observations and data analysis in Sec.~\ref{sec:obs}. The superburst properties, including the light curve, the detection of mHz QPOs and the spectral results, are shown in Sec.~\ref{sec:results}. We discuss the properties of the superburst, the mHz QPOs, the quenching time, and the recurrence of type I X-ray bursts in Sec.~\ref{sec:dis}.
% Several kinds of derivation from blackbody have been observed

\section{Observations} \label{sec:obs}

\subsection{MAXI}

The Gas Slit Camera \citep[GSC,][]{Mihara11} on board MAXI \citep{Matsuoka09}, mounted on the International Space Station (ISS), is an X-ray all-sky monitor that scans about 85\% of the whole sky every 92 minutes \citep{Sugizaki11}. MAXI/GSC has the capability of detecting intermediate duration bursts and superbursts in the 2--20 keV band \citep[see e.g.,][]{Serino16}. We downloaded all the MAXI/GSC observations of the second superburst in \aql\ between MJD 59130.126--59131.998 from the MAXI webpage,\footnote{\url{http://maxi.riken.jp/mxondem/}} including the light curves (2--6 keV), spectra (2--20 keV) and response files. The exposure of each spectrum is around 250 s.

\subsection{NICER}
%The Neutron star Interior Composition ExploreR (\nicer), launched on June 3, 2017, is an International Space Station (ISS) payload dedicated to (spectral) timing studies in the 0.2--12 keV band at an unprecedented time resolution of $\sim 100$ ns  \citep{arzoumanian2014}. \aql\ went into outburst in 2020. After launched in June 2017, the X-ray Timing Instrument (XTI) of \nicer\ is an ideal instrument to study type I X-ray burst, providing good energy resolution, high throughput and highly accurate time resolution in the 0.2--12 keV energy band. 
We analyzed the  \nicer\  ObsIDs 3050340143--59, which started before the superburst trigger (MJD 59129.6) and lasted to the end of the outburst (MJD 59157.2) with a total exposure time of 48.6 ks. We carried out standard data processing using HEASOFT version 6.28 and the NICER Data Analysis Software (NICERDAS). The default filtering criteria were applied to obtain the cleaned event data. Moreover, all event data are solar barycenterred by the tool {\tt barycorr}. We have checked the light curve in the energy range 10--12 keV for flaring background, but we did not find it. By using {\tt xselect}, we extracted 16 s light curves in the energy ranges 0.5--2.5 keV, 2.5--10 keV, and 0.5--10 keV, to show the hardness ratio (see Fig.~\ref{fig:lc}), 1 s  light curves in the energy range 0.5--10 keV to search the type I X-ray burst after the superburst (see Sec.~\ref{sec:lc}) and to explore mHz QPOs (see Sec.~\ref{subsec:osc}). We used the up-to-date ancillary response file (ARF) and response matrix file (RMF) to perform the spectral analysis (see Sec.~\ref{subsec:pers} and \ref{subsec:burst}).

\subsection{\hxmt}
The Insight Hard X-ray Modulation Telescope \citep[\hxmt,][]{hxmt}, launched on 2017 June 15, have three instruments, the Low Energy X-ray telescope \citep[LE, 1--15~keV, 384~cm$^2$;][]{hxmt-le}, the Medium Energy X-ray telescope \citep[ME, 5--30~keV, 952~cm$^2$,][]{hxmt-me}, and the High Energy X-ray telescope \citep[HE, 20--250~keV, 5100~cm$^2$,][]{hxmt-he}. \hxmt\ observed \aql\ during its 2020 outburst between MJD 59085.278--59143.741.
We employ only the LE data to search for type I X-ray bursts and mHz QPOs.  We analyze the data using the \hxmt\
Data Analysis Software package (HXMTDAS) version 2.04. The LE data are calibrated by using the scripts  {\tt lepical}. The good time intervals are selected from the scripts {\tt legtigen} with the standard criteria; that is, the Earth elevation angle (ELV) greater than 10$^{\circ}$, the cutoff rigidity (COR) higher than 8 GeV, the offset angle from the pointing source (ANG\_DIST) smaller than $0^{\circ}.04$, and the satellite located outside the South Atlantic Anomaly (SAA) region. The 1 s background substracted light curves after the superburst have been obtained with a total exposure time of 18.2 ks.

\subsection{\swift}
The Neil Gehrels Swift observatory satellite observed \aql\ during its 2020 outburst between 2020 August 20 (MJD 59081.68) and October 26 (MJD 59148.66). We only focus the data collected after  the peak of the superburst in \aql (i.e., MJD 59130.7) resulting in a total number of 17 observations. The
X-Ray Telescope \citep[XRT;][]{Burrows05} on board \swift\ operated mainly in windowed timing (WT) mode. The 1 s and 16 s XRT light curves in the energy range 0.3--10 keV were produced with the \swift\ XRT data products generator tool at the UK Swift Science Data Centre\footnote{\url{https://www.swift.ac.uk/user\_objects}}
\citep[see][for more details]{Evans07,Evans09}. The total exposure time is 8.7 ks.

\section{Results} \label{sec:results}
\subsection{Light curve}\label{sec:lc}

\begin{figure*}
\centering
\includegraphics[width=18cm]{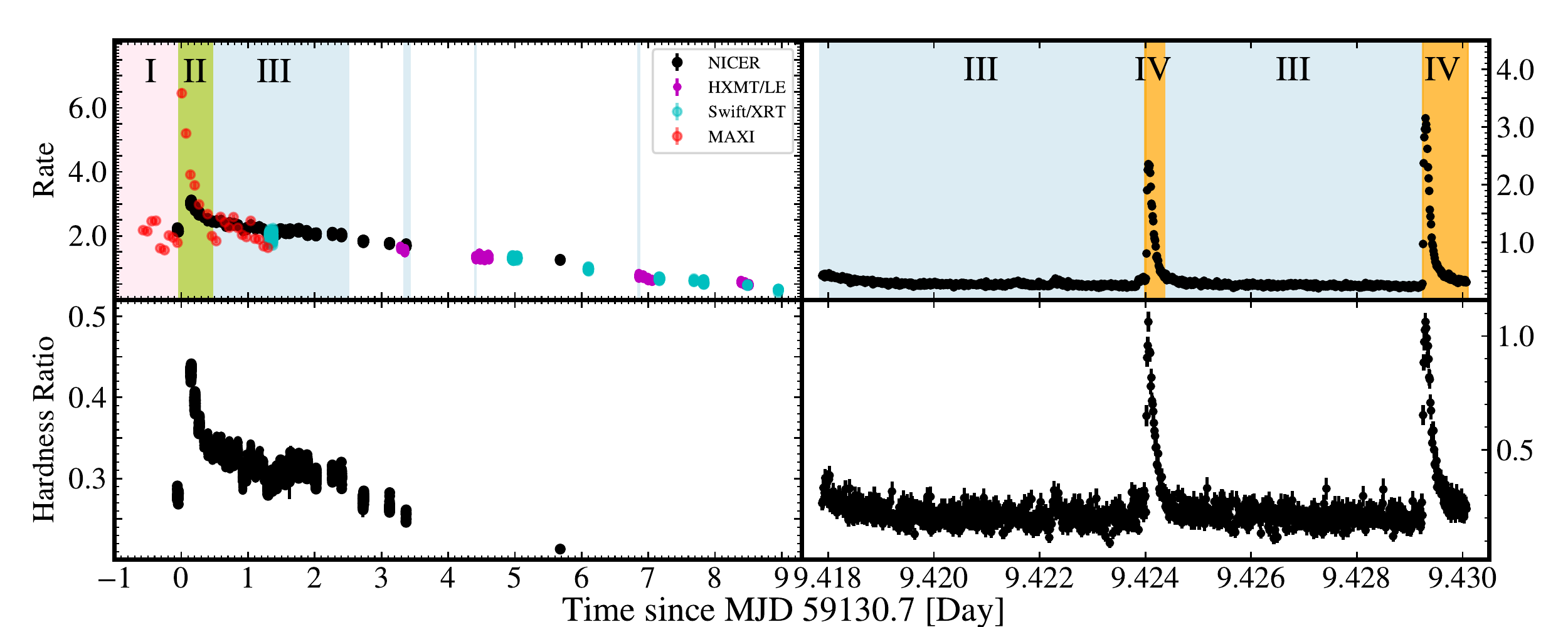}

\caption{Top left panel, the light curve of the Aql X-1 superburst from \nicer\ (16 s, 0.5--10 keV, , and indicated by black points), \swift/XRT (16 s, 0.3--10 keV, and cyan points), \hxmt/LE (16 s, 1--10 keV, and magenta points) and MAXI  ($\sim250$ s, 2--6 keV, and red points). Top right panel, \nicer\ observed (ObsID 305034015) two type I X-ray bursts from Aql X-1 since the superburst, and the 1 s  light curve in the energy range 0.5--10 keV is shown. The rate of \nicer, \swift/XRT and \hxmt/LE light curves are in units of $10^3~ \rm{c~s^{-1}}$, $80~ \rm{c~s^{-1}}$, and  $10^2~ \rm{c~s^{-1}}$, respectively. The MAXI data are in units of $ \rm{c~s^{-1}~cm^{-2}}$ and are multiplied by 2.5 to make a better visualization. It is not surprising that the time of the available \nicer\ and MAXI observations are well overlapped since both instruments are installed on ISS. The regions I, II, III and IV mark the time intervals of the pre-superburst, the superburst, the detection of mHz QPOs and the return of type I X-ray bursts, respectively. In the bottom left (right) panel, the hardness ratio between the \nicer\  2.5--10 keV and 0.5--2.5 keV 16 s (and 1 s) light curves are displayed.   \label{fig:lc}}
\end{figure*}

In the top left panel of Fig.~\ref{fig:lc}, we show the light curve of \aql\ from MAXI (binned with $\sim250$ s), \nicer\ (16 s), \swift/XRT (16 s) and \hxmt\ (16 s). We note that almost all \nicer, \swift/XRT and \hxmt observations showed multiple data gaps due to South Atlantic Anomaly passages, Earth occultations, and other constraints. Most observations have multiple, nearly contiguous segments, and each of them lasts $\sim$ 200--1800 s. The hardness ratio between 2--10 keV and 0.5--2 keV of \nicer\ observations are shown in the bottom left panel of Fig.~\ref{fig:lc}.  Before the superburst, the hardness ratio is around 0.28. During the decay phase, the hardness ratio decreased from 0.44 to 0.3 due to the cooling of the superburst emission. We also find that the post-superburst emission is harder than the pre-superburst. Assuming that the source accretion rate was not changing during the superburst period in a half-day, we propose the hardening just after the superburst is a consequence of the accretion interacting with the still very hot plasma around the NS; therefore, the source changed to a harder spectrum. We divide the light curves into four intervals: the pre-superburst (I), the superburst (II), the detection of mHz QPOs (III, see Sec.~\ref{subsec:osc}), and the return of type I X-ray bursts (IV), respectively.

After the superburst, we search for the return of the type I X-ray burst. We find two type I X-ray bursts 9.44 days after the peak of the superburst from \nicer, and none from \hxmt. On MJD 59140.518, 0.389 days after the second NICER burst, we find the third type I X-ray bursts from \swift/XRT ObsID 00033665175.  In the right panels of Fig.~\ref{fig:lc}, we plot the 0.5--10 keV light curve of two type I X-ray bursts from the ObsID 3050340150 and show the hardness ratio between 2.5--10 keV and 0.5--2.5 keV at 1 s resolution. During these two type I X-ray bursts, the hardness ratios evolve as the same trend of their light curves that is a fast rise and exponential decay, which are typical behaviors of a thermonuclear burst. We determine the start time for each burst is when the intensity reached to 10\% of the peak above the persistent intensity level. Then, we define the rise time as the time between the start of the burst and the time at which the intensity reached 90\% of
the peak burst intensity. For these two bursts, the rise time is around 2 s. The recurrence time of these two type I X-ray bursts is 7.6 minutes.

% \begin{figure}
% \includegraphics[width=8.5cm]{lc_3050340150.pdf}
% \caption{\nicer\ observed (ObsID 305034015) two type I X-ray bursts from Aql X-1 since the superburst. In the top panle, the 1 s  light curve in the energy range 0.5--10 keV is shown. The hardness ratio between the 1 s light curves in the energy ranges 2.5--10 keV and 0.5--2.5 keV are displayed in the bottom panel.  }
% \label{fig:Xray_burst}
% \end{figure}

\subsection{Search for burst oscillation and mHz QPOs} \label{subsec:osc}

We perform the $Z_1^2$-test statistics by using the code {\tt z\_n\_search} based on {\tt Stingray} \citep{Huppenkothen19} to search for burst oscillations in the frequency range 545--555 Hz. The cleaned event files in the 0.5--10 keV energy range are used by applying a moving window method, while the window sizes $T$ are 1, 2, and 4 s, with steps of $T/10$. No significant burst oscillation signals are observed. 

% \subsection{Search for mHz QPO} \label{subsec:mHz}

\begin{figure}
\includegraphics[width=8.5cm]{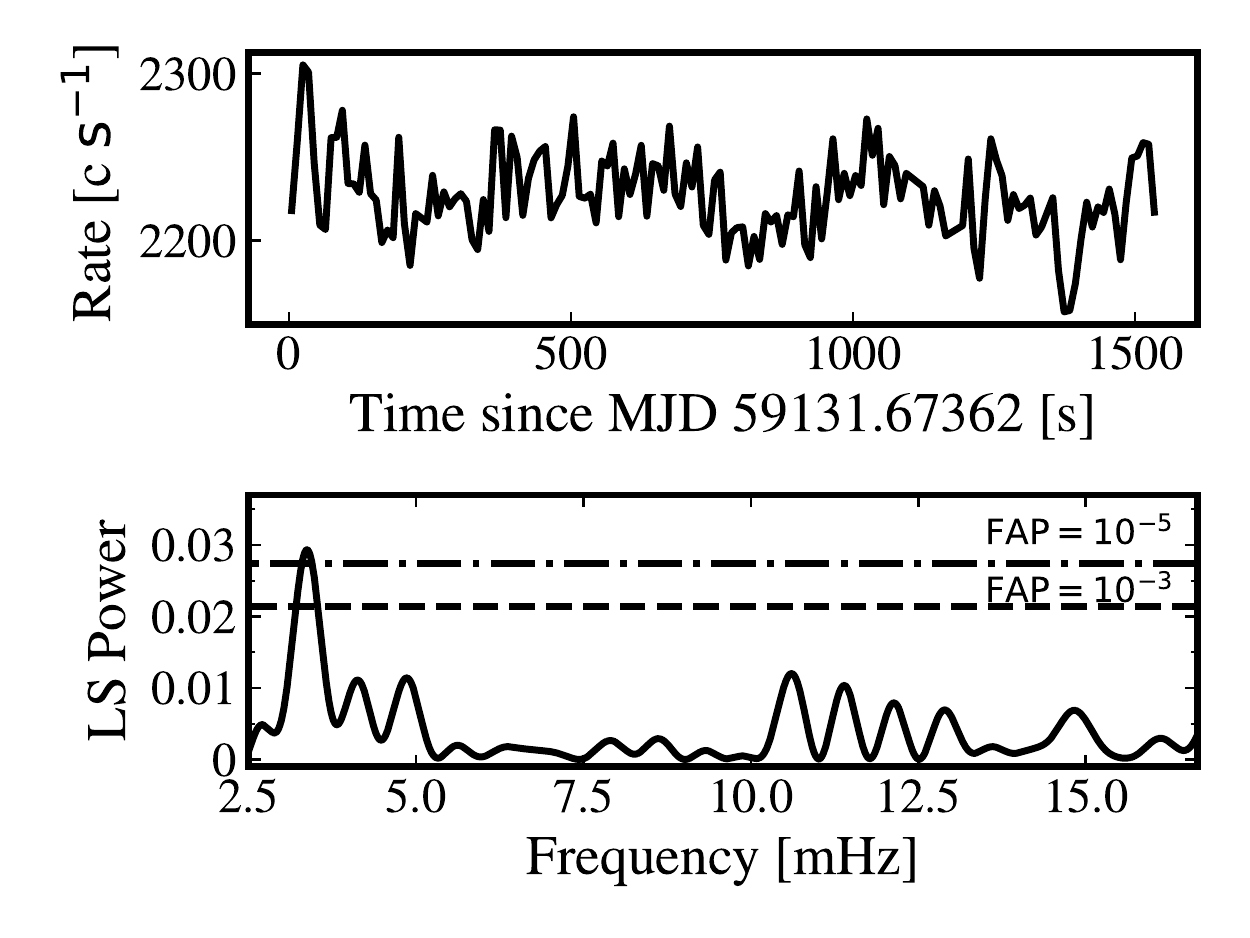}
\caption{Top panel, the 0.5--10 keV light curve of the segment NO. 11 in ObsID 3050340145 rebinned with 10 s. Bottom panel, the Lomb-Scargle periodogram of this segment. The maximum power occurs at 3.4 mHz with FAP $\sim2.1\times10^{-6}$. The dashed and dash-dotted lines represent the FAPs of $10^{-3}$ and $10^{-5}$, respectively.  }
\label{fig:QPO}
\end{figure}

We apply the Lomb--Scargle periodogram \citep{VanderPlas18} to all segments to search for mHz QPO from \nicer, \swift/XRT and \hxmt/LE, in the 0.5--10 keV, 0.3--10 keV, and 1--10 keV light curves, respectively. From \nicer\ observations, we find strong mHz QPO signals appearing from the fourth segment in ObsID 3050340145 to the first segment in ObsID 3050340147 (MJD 59131.2--59133.2), the only segment in ObsID 3050340148, and the intervals before the first type I X-ray burst and between the two bursts in ObsID 3050340150.  We also detect mHz QPO signals in ObsID 00033665163 from \swift/XRT, and ObsID P030403404101, ObsID P030403404201 and ObsID P030403404301 from \hxmt/LE. The frequencies of mHz QPO range from 2.7 to 11.3 mHz, corresponding to the timescales of 1.47--6.17 minutes. In particular, the time of the ObsID 00033665163 from \swift/XRT is partially overlapped with the second segment of the ObsID 3050340146 from \nicer, and the observed frequencies of mHz QPO are close, i.e., 4.8 mHz versus 5.0 mHz. The \hxmt/LE ObsID P030403404101 has been observed 0.022 days earlier than the \nicer\ ObsID 3050340148, the detected mHz QPOs from these observations are also similar.  In Table~\ref{table:QPO}, for each segment, we provide the duration, the start time, the frequency of mHz at the highest periodogram power, and its false alarm probability (FAP) with the assumption of Gaussian white noise \citep{Baluev08}. In Fig.~\ref{fig:QPO}, we show the light curve and the power of the eleventh segment in ObsID 3050340145 as an example.

%v\textbf{The second and third segments in ObsID 3050340147 showed comparable count rated compare with the previous and proceeding segments with detected mHz QPO, which means }

%=======================
\begin{table}%[h] 
{\small
\caption{The Detections of mHz QPO from \nicer, \swift/XRT and \hxmt/LE}. 
%\red{*It makes sense to add dates of the observations*}

\centering
\begin{tabular}{lccccc} 
\hline 
ObsID$^{\mathrm{a}}$  &  NO.$^{\mathrm{b}}$ & Time$^{\mathrm{c}}$ & $T$$^{\mathrm{d}}$ & $f$$^{\mathrm{e}}$ & FAP$^{\mathrm{f}}$ \\
 &  & Day & (ks)  &(mHz)   &   \\
\hline 
\noalign{\smallskip}  
\multicolumn{6}{c}{\nicer} \\
\hline 
3050340145    &  4  &  0.521   & 1.52 & 3.9   & $2.7\times10^{-9}  $    \\ 
              &  5  &  0.585   & 1.58 & 3.3   & $7.5\times10^{-7}  $    \\ 
              &  6  &   0.650  & 1.58 & 3.4   & $6.6\times10^{-17}  $    \\ 
              &  7  &  0.718  & 1.29 & 4.8   & $6.2\times10^{-5}   $    \\  
              &  8  &  0.779  & 1.61 & 2.9   & $5.0\times10^{-11}  $    \\  
              &  9  &  0.844  & 1.53 & 3.0   & $2.5\times10^{-5}   $    \\ 
              &  10 &  0.909  & 1.53 & 4.0   & $5.3\times10^{-5}   $     \\ 
              &  11 &  0.974  & 1.55 & 3.4   & $2.1\times10^{-6}   $    \\ 
              &  12 &   1.037  & 1.68 & 3.6   & $1.1\times10^{-10}  $    \\ 
              &  13 &   1.103  & 1.53 & 5.7   & $3.1\times10^{-4}   $    \\ 
              &  14 &   1.166  & 1.66 & 3.2   & $2.1\times10^{-4}   $    \\
              &  15 &   1.232  & 0.97 & 3.9   & $1.1\times10^{-8}   $    \\ 
3050340146    &  1  &   1.295 & 0.86 & 6.2   & 0.003                    \\ 
              &  2  &   1.359 & 0.97 & 5.0   & 0.003                    \\ 
              &  3  &   1.425 & 1.26 & 11.3  & $8.3\times10^{-7}  $     \\ 
              &  4  &   1.489 & 0.44 & 8.9   & 0.02                    \\  
              &  5  &   1.555 & 1.58 & 5.2   & 0.0015                  \\  
              &  6  &   1.625 & 1.60 & 8.9   & $2.6\times10^{-5}   $     \\ 
              &  7  &   1.748 & 1.62 & 3.7   & $1.5\times10^{-6}  $     \\ 
              &  8  &   1.876 & 1.73 & 2.7   & $2.1\times10^{-10}  $    \\ 
              &  9  &   2.005 & 1.75 & 6.2   & 0.009                    \\ 
              &  10 &   2.264 & 0.35 & 6.7   & 0.0002                   \\ 
3050340147    &  1  &   2.399 & 0.59 & 3.2   & $4.5\times10^{-14}  $    \\
3050340148 & 1 &  3.368    & 0.30 & 6.2 & $4.4\times10^{-8}  $ \\
3050340150    &  1  & 9.418 & 0.52 & 3.5   & $3.3\times10^{-24}  $    \\
              &  2  & 9.425 & 0.39 & 4.4   & $2.2\times10^{-6}  $    \\
              \hline 
\multicolumn{6}{c}{\swift/XRT} \\
\hline 
00033665163   &  2 & 1.370   & 0.75 & 4.8   & 0.001   \\
\hline 
\multicolumn{6}{c}{\hxmt/LE} \\
\hline 
P030403404101  &  2 & 3.346   & 0.47 & 6.9   & 0.005   \\
P030403404201 &  1  &4.407 & 0.75 & 2.7 & $1.5\times10^{-6}$   \\
P030403404301  &  1 & 6.854   & 1.0 & 3.2   & 0.02   \\
\noalign{\smallskip}  
\hline  
\end{tabular}  

$^{\mathrm{a}}$The ObsIDs 3050340145, 3050340146, 3050340147 have 15, 10, and 3 segments, respectively. 3050340150 has only one segment; however, we divide the data into two intervals, which locate before the first type I X-ray burst and between two bursts. 

\noindent$^{\mathrm{b}}$The number of the segment. 

\noindent$^{\mathrm{c}}$The start time of the segment since MJD 59130.7. 

\noindent$^{\mathrm{d}}$The duration of the segment. 

\noindent$^{\mathrm{e}}$The detected frequency of mHz QPO. 

\noindent$^{\mathrm{f}}$The false alarm probability at the highest periodogram peak.

\label{table:QPO} 
}
\end{table} 

\subsection{Pre-superburst and pre-burst emissions}\label{subsec:pers}

\begin{figure} [h]
\includegraphics[width=9.5cm]{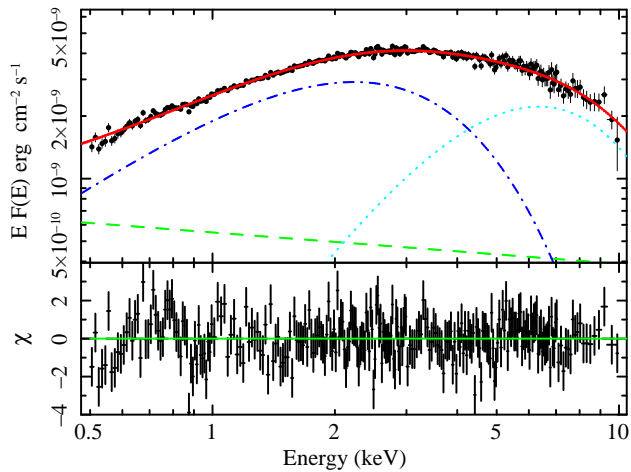}
\caption{The unfolded and unabsorbed pre-superburst spectrum of \aql\ in the 0.5--10 keV energy range. The best fit model has three components, a blackbody from the NS surface or boundary layer, a diskbb from the accretion disk and a power-law from the corona. The spectrum are rebinned for a better view. The residuals from the best fit are shown at the bottom. }
\label{fig:persistent}
\end{figure}

%, see Sec.~\ref{subsec:burst} for more details
We analyze the pre-superburst and two pre-burst spectra from \nicer\ observations, and one pre-burst spectrum from \swift/XRT observations, which are regarded as background during the superburst and the three type I X-ray bursts.  We extract the pre-superburst spectrum from the second segment in ObsID 3050340144, which has an exposure of 239 s and a mean count rate of $2.15\times10^3 ~\rm{c~s^{-1}}$. For the two pre-burst emissions from \nicer\ observations, the spectra are extracted from the 120 s data prior to the trigger of two type I X-ray bursts in ObsID 3050340150, resulting mean count rates of $2.5\times10^2 ~\rm{c~s^{-1}}$ and $2.4\times10^2 ~\rm{c~s^{-1}}$, respectively.  We generate the background spectra from the {\tt nibackgen3C50} tool \citep[v6; ][]{Remillard21} \footnote{\url{https://heasarc.gsfc.nasa.gov/docs/nicer/tools/nicer_bkg_est_tools.html}}, which have the count rates of $\sim$ 0.5 c/s. The photons from the source dominate these two persistent spectra in all energies. The spectra are grouped by a factor of 5.  For the pre-burst from \swift/XRT WT observations, the spectrum is extracted from the 1200 s data prior to the burst trigger by using \texttt{xselect} to filter a $70\arcsec$ circle centered on the source position, while the background spectrum is produced from an annulus with an inner and outer radii of $165\arcsec-200\arcsec$, respectively.  The spectrum is grouped to guarantee at least 25 photons in each channel. The arf file generated from \texttt{xrtmkarf}, and the rmf file \texttt{swxwt0s6\_20131212v015.rmf} are used for spectral fitting.  We fit the pre-superburst emission with a combination of a blackbody \citep[\texttt{bbodyrad} in XSPEC,][]{Arnaud96}, a multicolor blackbody component \citep[\texttt{diskbb},][]{Mitsuda84} and a power-law component (\texttt{powerlaw}), which are  absorbed from the interstellar matter using the model {\tt tbabs} \citep[see][]{Lin07}.  A goodness of fit per degree of freedom of $\chi^2_\nu=1.08(292)$  is obtained (see Fig.~\ref{fig:persistent}).  We note that the $\chi^2_\nu$ are close to other methods of grouping, i.e., by minimum signal-to-noise ratio. Thereafter, we report all fitted parameters at 1$\sigma$ confidence level.  The best-fitting blackbody temperature the inner accretion disk temperature, and the power-law index are $1.60\pm0.10$ keV,  $0.94\pm0.04$ keV and $2.09\pm0.38$, respectively.  We find that $N_{\rm  H}\sim5.14\times10^{21}~\rm{cm^{-2}}$  is well consistent with the values reported in \citet{Keek18} and \citet{Pinto13}. We use \texttt{cflux} to estimate the unabsorbed bolometric flux of $1.05\pm0.01\times10^{-8}~{\rm erg~cm^{-2}~s^{-1}}$ in the energy range 0.5--250 keV, corresponding to $\sim10.5\%F_{\rm Edd}$, if the Eddington limit $F_{\rm Edd}\approx10^{-7}~{\rm erg~cm^{-2}~s^{-1}}$ is adopted \citep{Li17,Keek18}. We adopt the same model \texttt{tbabs(powerlaw+gaussian)} to fit three pre-burst spectra. For the first pre-burst spectrum, the best-fitting power-law index, the energy and width of the gaussian component are $2.19\pm0.03$, $0.69\pm0.04$ keV and $0.12\pm0.02$ keV, respectively. The unabsorbed bolometric flux is $1.40\pm0.01\times10^{-9}~{\rm erg~cm^{-2}~s^{-1}}$, corresponding to $\sim1.4\%F_{\rm Edd}$. For the second pre-burst spectrum, we obtain that the power-law index, the energy and width of the gaussian component are $2.19\pm0.03$, $0.66\pm0.04$ keV and $0.16\pm0.03$ keV, respectively. The unabsorbed bolometric flux is $1.45\pm0.01\times10^{-9}~{\rm erg~cm^{-2}~s^{-1}}$, corresponding to $\sim1.5\%F_{\rm Edd}$. For the \swift/XRT pre-burst spectrum, the power-law index, the energy and width of the gaussian component are $2.06\pm0.05$, $0.44\pm0.15$ keV and $0.21\pm0.04$ keV, respectively. The unabsorbed bolometric flux is $1.17\pm0.02\times10^{-9}~{\rm erg~cm^{-2}~s^{-1}}$, corresponding to $\sim1.2\%F_{\rm Edd}$. The persistent emissions of the two type I X-ray bursts are only 13\% of the pre-superburst level, for the reason that the X-ray outburst of \aql\ was close to its end stage. 
%and $1\sigma$ uncertainties are presented in Table 1.

\subsection{Time resolved burst spectra}\label{subsec:burst}
\begin{figure}[h]
\includegraphics[width=8cm]{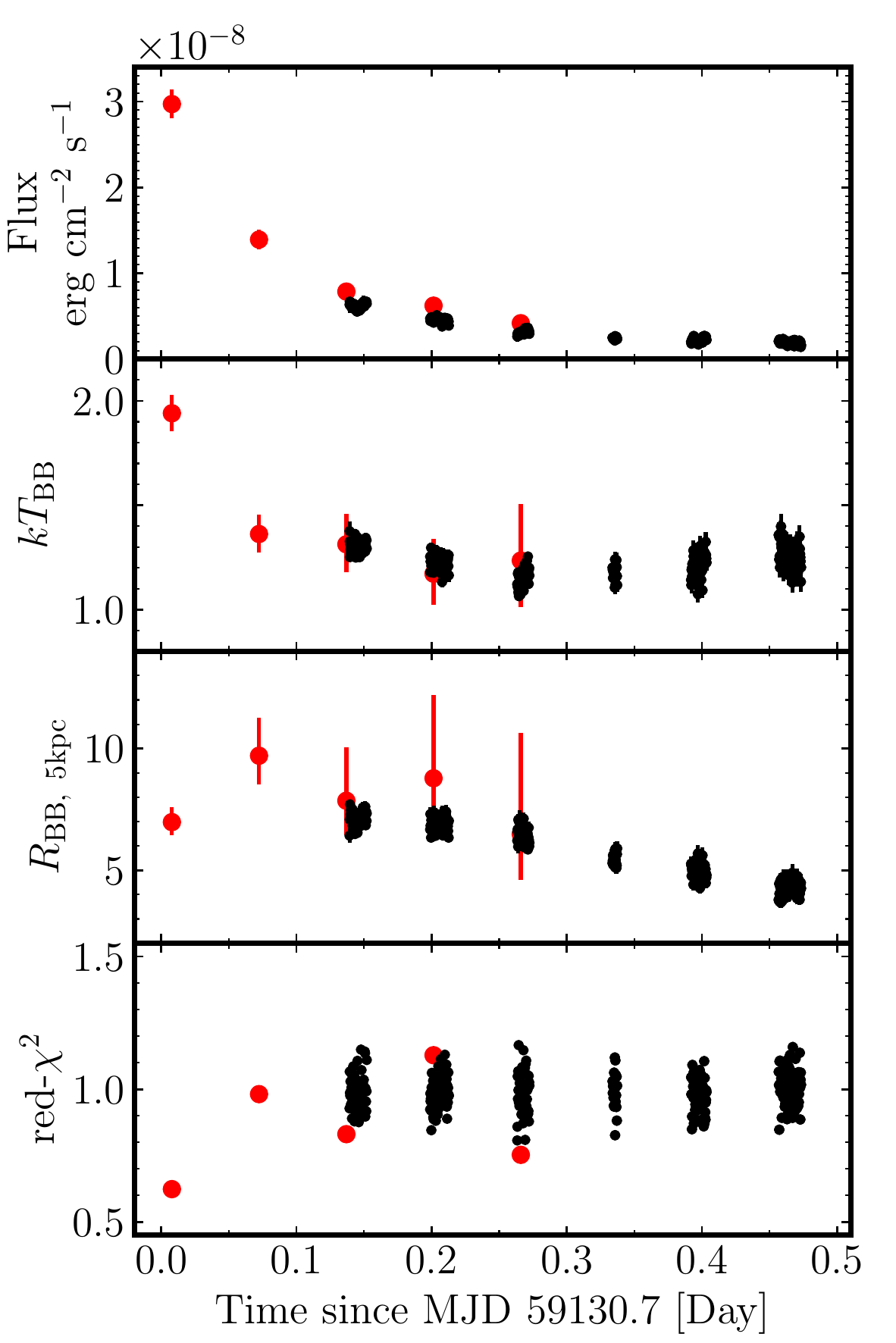}
\caption{The spectral parameters of the superburst in \aql\ evolve with time.   From top to bottom, the bolometric flux, the blackbody temperature and normalization, and the reduced-$\chi^2$ are shown. The red and black points are the data from MAXI and \nicer, respectively.}

\label{fig:burst}
\end{figure}

MAXI caught the early stages of the superburst from \aql, where \nicer\ performed sufficient observations since the later decay of the superburst.  For the \nicer\ observations, we extract the time resolved spectra with duration of 16 s.  The persistent emissions are considered as background during the burst emission. We group the burst spectra so that each bin contains a minimum of 20 counts. All \nicer\ and MAXI burst spectra are fitted by an absorbed blackbody model in the energy range 0.5--10 keV and 2--20 keV, respectively. The results are shown in Fig.~\ref{fig:burst}. The spectra during the superburst can be well fitted with an absorbed blackbody, with the temperature drops from $\sim$2 to 1.2 keV, and the blackbody radius of 3--10 km, assuming the distance of 5 kpc to \aql\ \citep{Rutledge01b}. We also fit three type I X-ray bursts after the superburst observed by \nicer\ and \swift/XRT. The peak fluxes of these three bursts are $2.24\pm0.33\times10^{-8}~{\rm erg~cm^{-2}~s^{-1}}$, $3.19\pm0.42\times10^{-8}~{\rm erg~cm^{-2}~s^{-1}}$, and $3.11\pm0.78\times10^{-8}~{\rm erg~cm^{-2}~s^{-1}}$, where the first burst is one of the faintest observed so far \citep{Galloway08,Galloway20}. We determine the e-folding decay times of these three bursts are $6.92\pm0.17$ s, $5.62\pm0.22$ s, and $9.00\pm0.61$ s, respectively.  Multiplied the peak flux by the e-folding time, we estimate the total burst fluences are $1.55\pm0.23\times10^{-7}~{\rm erg~cm^{-2}}$, $1.79\pm0.24\times10^{-7}~{\rm erg~cm^{-2}}$, and $2.80\pm0.73\times10^{-7}~{\rm erg~cm^{-2}}$, respectively. The hardness ratio variations of two \nicer\ type I X-ray bursts shown in Fig.~\ref{fig:lc} are a typical property of thermonuclear bursts caused by the blackbody evolution.
We do not find the signatures of enhanced persistent emission or reflection from the accretion disk.

\begin{figure} [h]
\includegraphics[width=8cm]{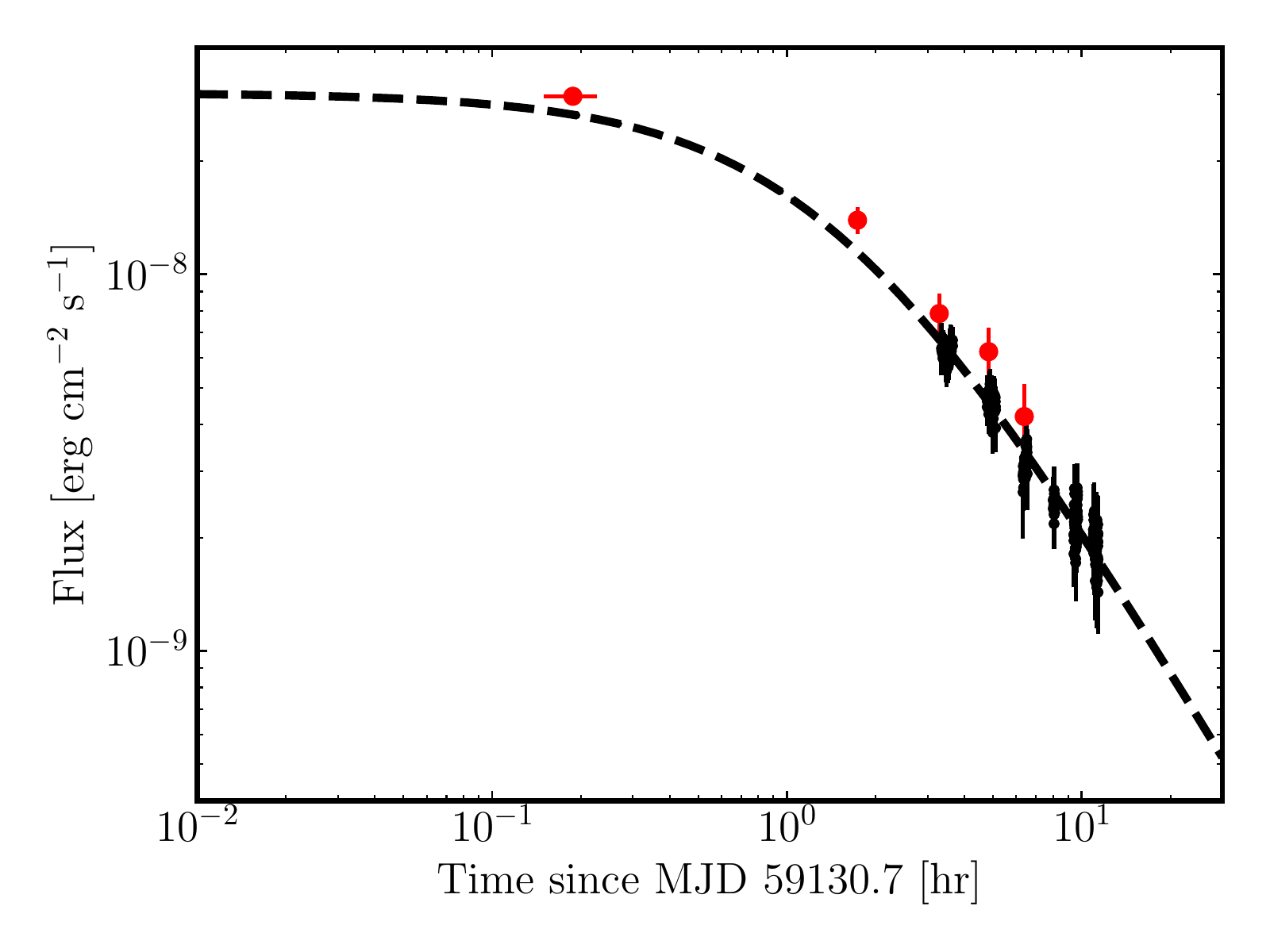}
\caption{The decay of the superburst fitted by the model proposed by \citet{Cumming04a}. The red and black points represent the data from MAXI and \nicer.}
\label{fig:decay}
\end{figure}

The decay of the superburst flux can be well described with the analytic expression by \citet[][see also \citealt{Cumming06}]{Cumming04a}, leading to constraints on the energy release per unit mass $E_{17}$ in
units of $10^{17}~\rm{erg~g^{-1}}$ and the ignition column depth $y_{12}$ in
units of $10^{12}~\rm{g~cm^{-2}}$. These two parameters depend on the time and flux of the superburst peak, which is unknown because the observations may not cover it \citep[see also][]{Serino16}. We assume that the peak flux started 1 hr earlier than the observed time, and we obtain $E_{17}\sim4.6$ and $y_{12}\sim2.7$, see Fig.~\ref{fig:decay}. We then calculate the total fluence of the superburst of $3.79\pm0.02\times10^{-4}~{\rm erg~cm^{-2}}$.  All the bursts parameters are reported in Table~\ref{table:burst}.

% %=======================
% \begin{table*}%[h] 
% \centering
% {\small
% \caption{The properties of the superburst and two type I X-ray bursts. 
% %\red{*It makes sense to add dates of the observations*}
% }
% \begin{tabular}{llll} 
% \hline 
%   & Superburst & First burst  & Second burst   \\
% \hline 
% \noalign{\smallskip}  
% Start time (MJD)   &  -   &  59140.12400 & 59140.12925     \\ 
% Peak time  (MJD)\tablenotemark{a}  &  59130.70783    &   59140.12406   & 59140.12931   \\ 
% Rise time (MJD)   &  -    &    5 s          & 5 s      \\
% $t_{\rm rec}$  &   7.2 yr        &   -    &      454 s    \\
% Peak flux   & $2.97\pm0.17$   & $2.24\pm0.33$ & $3.19\pm0.42$    \\
% ($10^{-8}~{\rm erg~cm^{-2}~s^{-1}}$)  &&& \\
% Persistent flux    &  $2.66\pm0.33$    & $0.080\pm0.007$&  $0.12\pm0.03$ \\          
% ($10^{-9}~{\rm erg~cm^{-2}~s^{-1}}$) & &  \\
%  $\tau$    &  $2.03\pm0.02$ hr     &  $8.08\pm0.15$ s & $7.20\pm0.15$ s      \\  
% \noalign{\smallskip}  
% \hline  
% \end{tabular}  
% \label{table:burst} 
% \tablefoot{ 
% \tablenotetext{a}{The peak of the superburst may be missed. Here, we only show the time at the maximum peak flux from MAXI.  }
% }
% }
% \end{table*} 

\begin{table*}%[h] 
\centering
{\small
\caption{The Properties of the Superburst and Three Type I X-Ray Bursts. 
%\red{*It makes sense to add dates of the observations*}
}
\begin{tabular}{lccccccc} 
\hline 
& Start Time  &  Rise Time & $t_{\rm rec}$ & Peak Flux & Persistent Flux & $\tau$ & $f_{\rm b}$$^{\mathrm{a}}$  \\ 
& MJD  & & &  $10^{-8}~{\rm erg~cm^{-2}~s^{-1}}$ & $10^{-9}~{\rm erg~cm^{-2}~s^{-1}}$  &  & $10^{-7}~{\rm erg~cm^{-2}}$\\
\hline 
\noalign{\smallskip}  
Superburst & 59130.68$\pm$0.02$^{\mathrm{b}}$  &  - & 7.2 yr & $2.97\pm0.17$ & $10.5\pm0.1$  & $2.03\pm0.02$ hr   & $3.79\pm0.02\times10^3$ \\
First burst$^{\mathrm{c}}$ &  59140.12400  &  $\sim$ 2 s & - & $2.24\pm0.33$ & $1.40\pm0.01$ &  $6.92\pm0.17$ s & $1.55\pm0.23$ \\
Second burst$^{\mathrm{c}}$ & 59140.12925  &  $\sim$ 2 s & 454 s & $3.19\pm0.42$  & $1.45\pm0.01$ &  $5.62\pm0.22$ s  &$1.79\pm0.24$ \\
Third burst$^{\mathrm{d}}$ & 59140.51856  &  $\sim$ 2 s & 0.389$^{\mathrm{e}}$ days & $3.11\pm0.78$  & $1.17\pm0.01$ &  $9.00\pm0.61$ s  &$2.80\pm0.73$ \\
\noalign{\smallskip}  
\hline  
\end{tabular}  
\label{table:burst} 
$^{\mathrm{a}}${Burst Fluence.} $^{\mathrm{b}}${The start time of the superburst was missed. We estimate the start time as the middle time of the gap before the superburst peak.  }  $^{\mathrm{c}}${Type I X-ray bursts from \nicer\ observations.}} $^{\mathrm{d}}${Type I X-ray burst from \swift/XRT observations.} $^{\mathrm{e}}${There are data gaps between the second and third bursts, and X-ray bursts may be missed. So, the recurrence time is only the upper limit. }

\end{table*}

\section{Discussion} \label{sec:dis}

% \subsection{Superburst properties}\label{subsec:sup_diss}
%

We first compare the two superbursts from the transient LMXB \aql.  MAXI detected two superbursts from \aql\ on MJD 56493.3 and 59130.7, indicating the recurrence time of $ \approx7.2$ yr. Assuming the distance of 5 kpc to \aql, the  bolometric flux and luminosity at the observed peak of count rate, the e-folding time, and the total energy of the first superburst are $2.0\pm0.3\times{\rm 10^{-8}~ erg~cm^{-2}~s^{-1}}$, $6.0\pm0.9\times{\rm 10^{37}~ erg~s^{-1}}$, 4.3 hr, and $9.3\pm1.4\times{\rm 10^{41}~ erg}$, respectively. However, in this work, we focus on the second superburst. The persistent emission of pre-superburst is well fitted with a model combination of blackbody, diskbb and power-law components. The bolometric flux and luminosity are $1.05\pm0.01\times10^{-8}~{\rm erg~cm^{-2}~s^{-1}}$ and $3.14\pm0.03\times10^{37}~{\rm erg~s^{-1}}$, respectively, corresponding to $\sim$10.5\% of the Eddington limit. The superburst lasted from MJD 59130.7 to 59131.2, and the flux evolved from $2.97\pm0.17\times10^{-8}$ to $1.43\pm0.32\times10^{-9}~{\rm erg~cm^{-2}~s^{-1}}$, see Table~\ref{table:burst}. From the decay of the superburst, we roughly estimate $E_{17}\sim4.6$ and $y_{12}\sim2.7$. The observed properties of two superbursts are comparable. 

After the superburst from \aql, for the first time, we observed the transition from superburst, via marginally stable burning, to unstable burning.  We then compare the observed quantities, including the quenching time, the recurrence of type I X-ray bursts after the superburst, and the frequencies of mHz QPO, with the simulations provided in \citet{Keek12}. \citet{Keek12} simulated the production and subsequent impacts of a superburst for a NS of $1.4M_\odot$ with a 10 km radius accreting composition by a solar, helium-rich and a mixture of carbon and iron (models H, He and C, respectively) at a constant rate of $\dot{M}=5.25\times10^{-9}M_{\odot}~\rm{yr^{-1}}$, corresponding to 30\% of the Eddington-limit rate. They found that the superburst occurs 1.28 yr after the start of accretion. For the case of \aql, the NS accretes the composition from the companion similar to the model H in \citet{Keek12}. However, there are few differences between the simulations and the observations in \aql. \aql\ is a transient source that has been observed X-ray outbursts recurring around one year, each of them lasting $\sim1$--6 months \citep{Degenaar19}.   Considering that the superburst recurrence time is $t_{\rm rec}\propto \dot{M^{-1}}$, we obtain $t_{\rm rec} \approx 1.28 (0.0854\dot{M}_{\rm Edd}/ 0.3\dot{M}_{\rm Edd})^{-1}(79/365)^{-1} {~\rm yr}\approx 20.8 {~\rm yr}$, if we take a typical value of the averaged accretion rate of $0.0854\dot{M}_{\rm Edd}$ in total 79 days from the 2013 outburst in \aql\ \citep{Ootes18}. Hence, the recurrence time of the superburst in \aql\ should be larger than the value we observed. However, the energy release from the inner crust may deduce the recurrence time due to frequent outbursts in \aql, similar to the case of GX 17+2 \citep[see also][]{Keek06}.
%, where $\dot{m}$ is the local accretion rate
%During an outburst, \aql\ emits at few percent of the Eddington flux, which is 10 times lower than the accretion rate used in \citet{Keek12}.From the pre-superburst to the time of three type I X-ray bursts, the persistent flux decreased from $\sim10.5\%F_{\rm Edd}$ to $\sim1.2\%F_{\rm Edd}$. Then, \aql\ evolved into the end stage of its 2020 outburst. 

When the NS cooling continues, with the quenching time of 1.1 days (model H) and 11.3 days (model He) after the superburst, type I X-ray bursts are returned.  We find that two weak type I X-ray bursts with a short recurrence time of 7.6 minutes appear 9.44 days after the superburst of \aql. We suggest that these two bursts are likely the first appearance of unstable burning from \aql\ since the superburst. There are data gaps, i.e., 0.625 days of total exposure with the time span of 8.924 days, between the end of superburst and the detection of type I X-ray bursts, which may miss bursts. However, we have 47 continuous segments lasting at least 500 s (\nicer, \hxmt/LE and \swift/XRT data), i.e., longer than the recurrence time of 7.6 minutes, so more bursts should be detected. If the recurrence time is  longer than 7.6 minutes, then we can estimate the probability of the number of bursts for a 0.625 days exposure in the total duration of 8.924 days. By applying a binomial distribution, the probability $P$ of detecting $n$ bursts out of $N$ is $P = C(N, n) \eta^n (1-\eta)^{(N-n)}$, where $\eta\approx0.07$ is the observation efficiency \citep[see][for more details]{Keek12}, and $n$ is zero in our case. So, we have possibilities higher than 90\% with the total number of bursts $N \leq 1$. There is a little chance that we missed potential bursts.  Moreover, the properties of the two type I X-ray bursts are consistent with the simulated burst resumption. We propose that this is the first accurate measurement of the quenching time, even though data gaps exist.  This is also the shortest quenching time in all superbursts so far, except for the upper limit 2.2 days from GX 17+2 that apparently has a higher accretion rate. If the simulated quenching time is corrected with the relation $t_{\rm quen}\propto\dot{M}^{-3/4}$ proposed by \citet{Cumming04a} (and note that the pre-superburst emits at $\sim 10.5\%\dot{M}_{\rm Edd}$ and drops to $\sim 1.4\%\dot{M}_{\rm Edd}$ till the return of type I X-ray bursts), we will obtain a quenching time, i.e., $\sim$ 3.7 days assuming an averaged accretion rate of $6\%\dot{M}_{\rm Edd}$. Moreover, the rapid rotation of \aql\ can delay the appearance of type I X-ray bursts, leading to a longer quenching time\citep{Piro07, Keek09, Keek12}. The simulated and observed quenching time can be matched. 
%However, we have many continuous segments (\hxmt/LE and \swift/XRT data) lasting longer than 7.6 minutes, if the recurrence time of bursts is 7.6 minutes, there should have been several more bursts detected.

%, close to the simulated value at 8.2 minutes. However, the accretion rate at the burst resumption of \aql\ is about 300 times lower than the simulation, the expected recurrence time should be $2.46\times10^3$ minutes, significantly larger than the observed result.We then obtain the lower limit of the recurrence time of 1.8 days, which is significantly larger than the recurrence time.

When the type I X-ray bursts return, we find recurrence time of 7.6 minutes. Now, we compare the observed and predicted recurrence time \citep[see e.g.,][]{Li18b}. The bolometric flux between the two type I X-ray bursts is $F_{\rm pers}=1.45\pm0.01\times10^{-10}~\rm{erg~cm^{-2}~s^{-1}}$. The luminosity is $L_{\rm pers}=4\pi D^2F_{\rm pers}\approx 4.3\times10^{36}~{\rm erg}$, assuming the distance of 5 kpc to \aql.  The local accretion rate per unit area onto the NS is $\dot{m}=L_{\rm pers}(1+z)(4\pi R^2(GM/R))^{-1}\approx2.4\times 10^3~{\rm g~cm^{-2}~s^{-1}}$, where $G$ is the gravitational constant, the gravitational redshift is $1+z=1.31$ for a NS with a mass of $1.4M_\odot$ and there is a radius of $R=10$ km. We then estimate the ignition column depth $y_{\rm ign}=4\pi f_{\rm b} D^2 (1+z)/(4\pi R^2Q_{\rm nuc})\approx1.0\times10^7~{\rm g~cm^{-2}}$, where $Q_{\rm nuc}\approx1.31+6.95\bar{X}-1.92\bar{X}^2~\rm{MeV~nucleon^{-1}}\approx5.23~\rm{MeV~nucleon^{-1}}$ is the nuclear energy produced for solar composition, i.e., $\bar{X}=0.7$ \citep{Goodwin19}. From the ignition column depth and local accretion rate, we calculate the recurrence time between the two bursts, $t_{\rm rec}=(y_{\rm ign}/\dot{m})(1+z)\approx90.3~{\rm minutes}$, significantly larger than the observed value. For a helium burst, the calculated recurrence time is four times longer. Therefore, the observed recurrence time is too short to accrete enough matter to trigger the second X-ray burst. Type I X-ray bursts with short recurrence times have also been observed in several sources, e.g.,  EXO 0748--676 \citep{Boirin07}, 4U 1636--536 \citep{Linares09}, and see also \citet{Keek10} and \citet{Galloway08}. We propose several possibilities that can reduce the recurrence time. First, the resumption of type I X-ray bursts may ignite at a shallow column depth if the NS envelope is sufficiently hot \citep[see e.g.,][]{Galloway21}. Second, the accreted matter on the NS surface is inhomogeneously distributed, causing a larger local accretion rate. Third, the first type I X-ray burst only depletes a considerably small fraction of accreting matter \citep{Keek17b}. Fourth, the rapid  rotation of \aql\ can induce mixing of freshly accreted fuel to explain the recurrence time on a timescale of 10 minutes \citep{Fujimoto93,Piro07, Keek09, Keek10,Galloway18}.

After the superburst, the NS cools down and transits from stable burning to marginally stable burning, causing oscillations in light curves. \citet{Heger07} reproduce the oscillation period to be the geometric mean of the accretion and thermal time scales $P_{\rm osc}=\sqrt{t_{\rm acc}t_{\rm therm}}\sim 100~\rm s$.  \citet{Keek12} obtain a similar time scale of 5 minutes (model H) and 20 minutes (model He).  We detect the mHz QPOs in the frequency range 2.7--11.3 mHz (time scales of 1.5--6.2 minutes) lasting 2 days after the superburst. In addition, the mHz QPOs are transiently appearing between $\sim$MJD 59133.2 and 59141.1, which was unpredicted in \citet{Keek12}. Since the occurrance of marginally stable burning sensitively depends on the NS surface temperature, the transient of mHz QPOs may present the temperature evolution of the combination effects of NS cooling and heating by accretion. The detected frequencies of mHz QPO are consistent with previous results \citep{Revnivtsev01,Mancuso21}.  Before the first type I X-ray burst in ObsID 3050340150, a 3.5 mHz QPO appeares again with FAP$\sim3.3\times10^{-24}$. Moreover, we also observe a 4.4 mHz QPO between two bursts, which is unexpected from simulations. We propose that the first type I X-ray burst heated the NS envelope, causing the appearance of marginally stable burning before the second type I X-ray burst. This effect should be verified and taken into account in future simulations.

It has been four years since the launching of \nicer. It is possible that \nicer\ has been detected more superbursts  besides of the one discussed in this article. There are many signatures of superburst, including the enhanced emission, the hardness ratio, and the appearance of mHz QPOs; these signatures can be used to determine the presence of superbursts even if the observations only partially cover the source. It is worthwhile to perform a dedicated search for superbursts from \nicer\ data set, accompanying with other instruments, i.e., MAXI and \hxmt. 

% \subsection{The mHz QPOs}\label{subsec:QPO_diss}

% \section{Conclusion}

\acknowledgments

We thank the anonymous referee for comments and suggestions that improved our manuscript. Z.L. was supported by National Natural Science Foundation of China  (U1938107, U1838111).  This research has made use of data obtained from the High Energy Astrophysics Science Archive Research Center (HEASARC), provided by NASA Goddard Space Flight Center,  from the HXMT mission, a project funded by the China National Space Administration (CNSA) and the Chinese Academy of Sciences (CAS), and also from MAXI provided by RIKEN, JAXA, and the MAXI team.

% \facilities{\nicer, MAXI, \hxmt}

% %% Similar to \facility{}, there is the optional \software command to allow 
% %% authors a place to specify which programs were used during the creation of 
% %% the manuscript. Authors should list each code and include either a
% %% citation or url to the code inside ()s when available.

% \software{astropy \citep{Astropy},  
%           HEADAS \citep{HEAsoft}, Stingray \citep{Huppenkothen19} 
%           }

% \vspace{25mm}

%% Appendix material should be preceded with a single \appendix command.
%% There should be a \section command for each appendix. Mark appendix
%% subsections with the same markup you use in the main body of the paper.

%% Each Appendix (indicated with \section) will be lettered A, B, C, etc.
%% The equation counter will reset when it encounters the \appendix
%% command and will number appendix equations (A1), (A2), etc. The
%% Figure and Table counter will not reset.

\bibliography{ms_AqlX1.bbl}{}

\begin{thebibliography}{}
\expandafter\ifx\csname natexlab\endcsname\relax\def\natexlab#1{#1}\fi
\providecommand{\url}[1]{\href{#1}{#1}}
\providecommand{\dodoi}[1]{doi:~\href{http://doi.org/#1}{\nolinkurl{#1}}}
\providecommand{\doeprint}[1]{\href{http://ascl.net/#1}{\nolinkurl{http://ascl.net/#1}}}
\providecommand{\doarXiv}[1]{\href{https://arxiv.org/abs/#1}{\nolinkurl{https://arxiv.org/abs/#1}}}

\bibitem[{{Arnaud}(1996)}]{Arnaud96}
{Arnaud}, K.~A. 1996, in Astronomical Society of the Pacific Conference Series,
  Vol. 101, Astronomical Data Analysis Software and Systems V, ed. G.~H.
  {Jacoby} \& J.~{Barnes}, 17

\bibitem[{{Baluev}(2008)}]{Baluev08}
{Baluev}, R.~V. 2008, \mnras, 385, 1279,
  \dodoi{10.1111/j.1365-2966.2008.12689.x}

\bibitem[{{Boirin} {et~al.}(2007){Boirin}, {Keek}, {M{\'e}ndez}, {Cumming},
  {in't Zand}, {Cottam}, {Paerels}, \& {Lewin}}]{Boirin07}
{Boirin}, L., {Keek}, L., {M{\'e}ndez}, M., {et~al.} 2007, \aap, 465, 559,
  \dodoi{10.1051/0004-6361:20066204}

\bibitem[{{Burrows} {et~al.}(2005){Burrows}, {Hill}, {Nousek}, {Kennea},
  {Wells}, {Osborne}, {Abbey}, {Beardmore}, {Mukerjee}, {Short}, {Chincarini},
  {Campana}, {Citterio}, {Moretti}, {Pagani}, {Tagliaferri}, {Giommi},
  {Capalbi}, {Tamburelli}, {Angelini}, {Cusumano}, {Br{\"a}uninger}, {Burkert},
  \& {Hartner}}]{Burrows05}
{Burrows}, D.~N., {Hill}, J.~E., {Nousek}, J.~A., {et~al.} 2005, \ssr, 120,
  165, \dodoi{10.1007/s11214-005-5097-2}

\bibitem[{{Callanan} {et~al.}(1999){Callanan}, {Filippenko}, \&
  {Garcia}}]{Callanan99}
{Callanan}, P.~J., {Filippenko}, A.~V., \& {Garcia}, M.~R. 1999, \iaucirc, 7086

\bibitem[{{Campana} {et~al.}(2013){Campana}, {Coti Zelati}, \&
  {D'Avanzo}}]{Campana13}
{Campana}, S., {Coti Zelati}, F., \& {D'Avanzo}, P. 2013, \mnras, 432, 1695,
  \dodoi{10.1093/mnras/stt604}

\bibitem[{{Campana} {et~al.}(1998){Campana}, {Stella}, {Mereghetti}, {Colpi},
  {Tavani}, {Ricci}, {Dal Fiume}, \& {Belloni}}]{Campana98}
{Campana}, S., {Stella}, L., {Mereghetti}, S., {et~al.} 1998, \apjl, 499, L65,
  \dodoi{10.1086/311357}

\bibitem[{{Cao} {et~al.}(2020){Cao}, {Jiang}, {Meng}, {Zhang}, {Luo}, {Yang},
  {Zhang}, {Gu}, {Sun}, {Liu}, {Yang}, {Li}, {Tan}, {Liu}, {Du}, {Lu}, {Xu},
  {Guan}, {Zhang}, {Wang}, {Li}, {Zhang}, {Wen}, {Qu}, {Song}, {Li}, {Ge},
  {Zhou}, {Xiong}, {Zhang}, {Zhang}, {Cheng}, {Zhang}, {Li}, {Liang}, {Gao},
  {Yang}, {Liu}, {Liu}, {Yang}, \& {Zhang}}]{hxmt-me}
{Cao}, X., {Jiang}, W., {Meng}, B., {et~al.} 2020, Science China Physics,
  Mechanics, and Astronomy, 63, 249504, \dodoi{10.1007/s11433-019-1506-1}

\bibitem[{{Casella} {et~al.}(2008){Casella}, {Altamirano}, {Patruno},
  {Wijnands}, \& {van der Klis}}]{Casella08}
{Casella}, P., {Altamirano}, D., {Patruno}, A., {Wijnands}, R., \& {van der
  Klis}, M. 2008, \apjl, 674, L41, \dodoi{10.1086/528982}

\bibitem[{{Chen} {et~al.}(2020){Chen}, {Cui}, {Li}, {Wang}, {Xu}, {Lu}, {Wang},
  {Chen}, {Han}, {Hu}, {Zhang}, {Huo}, {Yang}, {Li}, {Lu}, {Zhang}, {Li},
  {Zhang}, {Xiong}, {Zhang}, {Xue}, {Zhao}, {Zhu}, {Zhu}, {Liu}, {Yang}, \&
  {Zhang}}]{hxmt-le}
{Chen}, Y., {Cui}, W., {Li}, W., {et~al.} 2020, Science China Physics,
  Mechanics, and Astronomy, 63, 249505, \dodoi{10.1007/s11433-019-1469-5}

\bibitem[{{Chevalier} {et~al.}(1999){Chevalier}, {Ilovaisky}, {Leisy}, \&
  {Patat}}]{Chevalier99}
{Chevalier}, C., {Ilovaisky}, S.~A., {Leisy}, P., \& {Patat}, F. 1999, \aap,
  347, L51

\bibitem[{{Cumming} \& {Bildsten}(2001)}]{Cumming01}
{Cumming}, A., \& {Bildsten}, L. 2001, \apjl, 559, L127, \dodoi{10.1086/323937}

\bibitem[{{Cumming} \& {Macbeth}(2004)}]{Cumming04a}
{Cumming}, A., \& {Macbeth}, J. 2004, \apjl, 603, L37, \dodoi{10.1086/382873}

\bibitem[{{Cumming} {et~al.}(2006){Cumming}, {Macbeth}, {in 't Zand}, \&
  {Page}}]{Cumming06}
{Cumming}, A., {Macbeth}, J., {in 't Zand}, J.~J.~M., \& {Page}, D. 2006, \apj,
  646, 429, \dodoi{10.1086/504698}

\bibitem[{{Degenaar} {et~al.}(2019){Degenaar}, {Ootes}, {Page}, {Wijnands},
  {Parikh}, {Homan}, {Cackett}, {Miller}, {Altamirano}, \&
  {Linares}}]{Degenaar19}
{Degenaar}, N., {Ootes}, L.~S., {Page}, D., {et~al.} 2019, \mnras, 488, 4477,
  \dodoi{10.1093/mnras/stz1963}

\bibitem[{{Evans} {et~al.}(2007){Evans}, {Beardmore}, {Page}, {Tyler},
  {Osborne}, {Goad}, {O'Brien}, {Vetere}, {Racusin}, {Morris}, {Burrows},
  {Capalbi}, {Perri}, {Gehrels}, \& {Romano}}]{Evans07}
{Evans}, P.~A., {Beardmore}, A.~P., {Page}, K.~L., {et~al.} 2007, \aap, 469,
  379, \dodoi{10.1051/0004-6361:20077530}

\bibitem[{{Evans} {et~al.}(2009){Evans}, {Beardmore}, {Page}, {Osborne},
  {O'Brien}, {Willingale}, {Starling}, {Burrows}, {Godet}, {Vetere}, {Racusin},
  {Goad}, {Wiersema}, {Angelini}, {Capalbi}, {Chincarini}, {Gehrels}, {Kennea},
  {Margutti}, {Morris}, {Mountford}, {Pagani}, {Perri}, {Romano}, \&
  {Tanvir}}]{Evans09}
---. 2009, \mnras, 397, 1177, \dodoi{10.1111/j.1365-2966.2009.14913.x}

\bibitem[{{Falanga} {et~al.}(2008){Falanga}, {Chenevez}, {Cumming}, {Kuulkers},
  {Trap}, \& {Goldwurm}}]{Falanga08}
{Falanga}, M., {Chenevez}, J., {Cumming}, A., {et~al.} 2008, \aap, 484, 43,
  \dodoi{10.1051/0004-6361:20078982}

\bibitem[{{Fujimoto}(1993)}]{Fujimoto93}
{Fujimoto}, M.~Y. 1993, \apj, 419, 768, \dodoi{10.1086/173528}

\bibitem[{{Galloway} \& {Keek}(2021)}]{Galloway21}
{Galloway}, D.~K., \& {Keek}, L. 2021, Astrophysics and Space Science Library,
  461, 209, \dodoi{10.1007/978-3-662-62110-3\_5}

\bibitem[{{Galloway} {et~al.}(2008){Galloway}, {Muno}, {Hartman}, {Psaltis}, \&
  {Chakrabarty}}]{Galloway08}
{Galloway}, D.~K., {Muno}, M.~P., {Hartman}, J.~M., {Psaltis}, D., \&
  {Chakrabarty}, D. 2008, \apjs, 179, 360, \dodoi{10.1086/592044}

\bibitem[{{Galloway} {et~al.}(2018){Galloway}, {in 't Zand}, {Chenevez},
  {Keek}, {Sanchez-Fernandez}, {Worpel}, {Lampe}, {Kuulkers}, {Watts}, {Ootes},
  \& {MINBAR Collaboration}}]{Galloway18}
{Galloway}, D.~K., {in 't Zand}, J. J.~M., {Chenevez}, J., {et~al.} 2018,
  \apjl, 857, L24, \dodoi{10.3847/2041-8213/aabd32}

\bibitem[{{Galloway} {et~al.}(2020){Galloway}, {in't Zand}, {Chenevez},
  {W{\"o}rpel}, {Keek}, {Ootes}, {Watts}, {Gisler}, {Sanchez-Fernandez}, \&
  {Kuulkers}}]{Galloway20}
{Galloway}, D.~K., {in't Zand}, J., {Chenevez}, J., {et~al.} 2020, \apjs, 249,
  32, \dodoi{10.3847/1538-4365/ab9f2e}

\bibitem[{{Goodwin} {et~al.}(2019){Goodwin}, {Heger}, \&
  {Galloway}}]{Goodwin19}
{Goodwin}, A.~J., {Heger}, A., \& {Galloway}, D.~K. 2019, \apj, 870, 64,
  \dodoi{10.3847/1538-4357/aaeed2}

\bibitem[{{Heger} {et~al.}(2007){Heger}, {Cumming}, \& {Woosley}}]{Heger07}
{Heger}, A., {Cumming}, A., \& {Woosley}, S.~E. 2007, \apj, 665, 1311,
  \dodoi{10.1086/517491}

\bibitem[{{Huppenkothen} {et~al.}(2019){Huppenkothen}, {Bachetti}, {Stevens},
  {Migliari}, {Balm}, {Hammad}, {Khan}, {Mishra}, {Rashid}, {Sharma}, {Martinez
  Ribeiro}, \& {Valles Blanco}}]{Huppenkothen19}
{Huppenkothen}, D., {Bachetti}, M., {Stevens}, A.~L., {et~al.} 2019, \apj, 881,
  39, \dodoi{10.3847/1538-4357/ab258d}

\bibitem[{{in't Zand}(2017)}]{Zand17}
{in't Zand}, J. 2017, in 7 years of MAXI: monitoring X-ray Transients, ed.
  M.~{Serino}, M.~{Shidatsu}, W.~{Iwakiri}, \& T.~{Mihara}, 121.
\newblock \doarXiv{1702.04899}

\bibitem[{{Iwakiri} {et~al.}(2020){Iwakiri}, {Serino}, {Negoro}, {Nakajima},
  {Aoki}, {Kobayashi}, {Takagi}, {Asakuram K}, {Seino}, {Mihara}, {Guo},
  {Zhou}, {Tamagawa}, {Matsuoka}, {Sakamoto}, {Sugita}, {Nishida}, {Komachi},
  {Yoshida}, {Tsuboi}, {Sasaki}, {Kawai}, {Okamoto}, {Kitakoga}, {Shidatsu},
  {Kawai}, {Adachi}, {Niwano}, {Nakahira}, {Sugawara}, {Ueno}, {Tomida},
  {Ishikawa}, {Tominaga}, {Nagatsuka}, {Ueda}, {Yamada}, {Ogawa}, {Setoguchi},
  {Yoshitake}, {Goto}, {Uematsu}, {Tsunemi}, {Yamauchi}, {Kurogi}, {Miike},
  {Kawamuro}, {Yamaoka}, {Kawakubo}, \& {Sugizaki}}]{Iwakiri20}
{Iwakiri}, W., {Serino}, M., {Negoro}, H., {et~al.} 2020, The Astronomer's
  Telegram, 14079, 1

\bibitem[{{Keek} {et~al.}(2010){Keek}, {Galloway}, {in't Zand}, \&
  {Heger}}]{Keek10}
{Keek}, L., {Galloway}, D.~K., {in't Zand}, J.~J.~M., \& {Heger}, A. 2010,
  \apj, 718, 292, \dodoi{10.1088/0004-637X/718/1/292}

\bibitem[{{Keek} \& {Heger}(2011)}]{Keek11}
{Keek}, L., \& {Heger}, A. 2011, \apj, 743, 189,
  \dodoi{10.1088/0004-637X/743/2/189}

\bibitem[{{Keek} \& {Heger}(2017)}]{Keek17b}
---. 2017, \apj, 842, 113, \dodoi{10.3847/1538-4357/aa7748}

\bibitem[{{Keek} {et~al.}(2012){Keek}, {Heger}, \& {in't Zand}}]{Keek12}
{Keek}, L., {Heger}, A., \& {in't Zand}, J.~J.~M. 2012, \apj, 752, 150,
  \dodoi{10.1088/0004-637X/752/2/150}

\bibitem[{{Keek} {et~al.}(2006){Keek}, {in't Zand}, \& {Cumming}}]{Keek06}
{Keek}, L., {in't Zand}, J.~J.~M., \& {Cumming}, A. 2006, \aap, 455, 1031,
  \dodoi{10.1051/0004-6361:20064884}

\bibitem[{{Keek} {et~al.}(2009){Keek}, {Langer}, \& {in't Zand}}]{Keek09}
{Keek}, L., {Langer}, N., \& {in't Zand}, J.~J.~M. 2009, \aap, 502, 871,
  \dodoi{10.1051/0004-6361/200911619}

\bibitem[{{Keek} {et~al.}(2018){Keek}, {Arzoumanian}, {Bult}, {Cackett},
  {Chakrabarty}, {Chenevez}, {Fabian}, {Gendreau}, {Guillot}, {G{\"u}ver},
  {Homan}, {Jaisawal}, {Lamb}, {Ludlam}, {Mahmoodifar}, {Markwardt}, {Miller},
  {Prigozhin}, {Soong}, {Strohmayer}, \& {Wolff}}]{Keek18}
{Keek}, L., {Arzoumanian}, Z., {Bult}, P., {et~al.} 2018, \apjl, 855, L4,
  \dodoi{10.3847/2041-8213/aab104}

\bibitem[{{Lewin} {et~al.}(1993){Lewin}, {van Paradijs}, \& {Taam}}]{Lewin93}
{Lewin}, W.~H.~G., {van Paradijs}, J., \& {Taam}, R.~E. 1993, \ssr, 62, 223,
  \dodoi{10.1007/BF00196124}

\bibitem[{{Li} {et~al.}(2017){Li}, {Falanga}, {Chen}, {Qu}, \& {Xu}}]{Li17}
{Li}, Z., {Falanga}, M., {Chen}, L., {Qu}, J., \& {Xu}, R. 2017, \apj, 845, 8,
  \dodoi{10.3847/1538-4357/aa7d0b}

\bibitem[{{Li} {et~al.}(2018){Li}, {De Falco}, {Falanga}, {Bozzo}, {Kuiper},
  {Poutanen}, {Cumming}, {Galloway}, \& {Zhang}}]{Li18b}
{Li}, Z., {De Falco}, V., {Falanga}, M., {et~al.} 2018, \aap, 620, A114,
  \dodoi{10.1051/0004-6361/201833857}

\bibitem[{{Lin} {et~al.}(2007){Lin}, {Remillard}, \& {Homan}}]{Lin07}
{Lin}, D., {Remillard}, R.~A., \& {Homan}, J. 2007, \apj, 667, 1073,
  \dodoi{10.1086/521181}

\bibitem[{{Linares} {et~al.}(2009){Linares}, {Watts}, {Altamirano}, {Patruno},
  {Casella}, {Rea}, {Soleri}, {van der Klis}, {Wijnands}, {Belloni}, {Homan},
  \& {Mendez}}]{Linares09}
{Linares}, M., {Watts}, A., {Altamirano}, D., {et~al.} 2009, The Astronomer's
  Telegram, 1979, 1

\bibitem[{{Liu} {et~al.}(2020){Liu}, {Zhang}, {Li}, {Lu}, {Chang}, {Li},
  {Zhang}, {Jin}, {Yu}, {Zhang}, {Fu}, {Chen}, {Ji}, {Xu}, {Deng}, {Shang},
  {Liu}, {Lu}, {Zhang}, {Dong}, {Li}, {Wu}, {Li}, {Wang}, {Wu}, {Zhang},
  {Zhang}, {Xiong}, {Liu}, {Zhang}, {Liu}, {Yang}, \& {Zhang}}]{hxmt-he}
{Liu}, C., {Zhang}, Y., {Li}, X., {et~al.} 2020, Science China Physics,
  Mechanics, and Astronomy, 63, 249503, \dodoi{10.1007/s11433-019-1486-x}

\bibitem[{{Mancuso} {et~al.}(2021){Mancuso}, {Altamirano}, {M{\'e}ndez}, {Lyu},
  \& {Combi}}]{Mancuso21}
{Mancuso}, G.~C., {Altamirano}, D., {M{\'e}ndez}, M., {Lyu}, M., \& {Combi},
  J.~A. 2021, \mnras, 502, 1856, \dodoi{10.1093/mnras/stab159}

\bibitem[{{Mata S{\'a}nchez} {et~al.}(2017){Mata S{\'a}nchez},
  {Mu{\~n}oz-Darias}, {Casares}, \& {Jim{\'e}nez-Ibarra}}]{Mata17}
{Mata S{\'a}nchez}, D., {Mu{\~n}oz-Darias}, T., {Casares}, J., \&
  {Jim{\'e}nez-Ibarra}, F. 2017, \mnras, 464, L41,
  \dodoi{10.1093/mnrasl/slw172}

\bibitem[{{Matsuoka} {et~al.}(2009){Matsuoka}, {Kawasaki}, {Ueno}, {Tomida},
  {Kohama}, {Suzuki}, {Adachi}, {Ishikawa}, {Mihara}, {Sugizaki}, {Isobe},
  {Nakagawa}, {Tsunemi}, {Miyata}, {Kawai}, {Kataoka}, {Morii}, {Yoshida},
  {Negoro}, {Nakajima}, {Ueda}, {Chujo}, {Yamaoka}, {Yamazaki}, {Nakahira},
  {You}, {Ishiwata}, {Miyoshi}, {Eguchi}, {Hiroi}, {Katayama}, \&
  {Ebisawa}}]{Matsuoka09}
{Matsuoka}, M., {Kawasaki}, K., {Ueno}, S., {et~al.} 2009, \pasj, 61, 999,
  \dodoi{10.1093/pasj/61.5.999}

\bibitem[{{Mihara} {et~al.}(2011){Mihara}, {Nakajima}, {Sugizaki}, {Serino},
  {Matsuoka}, {Kohama}, {Kawasaki}, {Tomida}, {Ueno}, {Kawai}, {Kataoka},
  {Morii}, {Yoshida}, {Yamaoka}, {Nakahira}, {Negoro}, {Isobe}, {Yamauchi}, \&
  {Sakurai}}]{Mihara11}
{Mihara}, T., {Nakajima}, M., {Sugizaki}, M., {et~al.} 2011, \pasj, 63, S623,
  \dodoi{10.1093/pasj/63.sp3.S623}

\bibitem[{{Mitsuda} {et~al.}(1984){Mitsuda}, {Inoue}, {Koyama}, {Makishima},
  {Matsuoka}, {Ogawara}, {Shibazaki}, {Suzuki}, {Tanaka}, \&
  {Hirano}}]{Mitsuda84}
{Mitsuda}, K., {Inoue}, H., {Koyama}, K., {et~al.} 1984, \pasj, 36, 741

\bibitem[{{Ootes} {et~al.}(2018){Ootes}, {Wijnands}, {Page}, \&
  {Degenaar}}]{Ootes18}
{Ootes}, L.~S., {Wijnands}, R., {Page}, D., \& {Degenaar}, N. 2018, \mnras,
  477, 2900, \dodoi{10.1093/mnras/sty825}

\bibitem[{{Pinto} {et~al.}(2013){Pinto}, {Kaastra}, {Costantini}, \& {de
  Vries}}]{Pinto13}
{Pinto}, C., {Kaastra}, J.~S., {Costantini}, E., \& {de Vries}, C. 2013, \aap,
  551, A25, \dodoi{10.1051/0004-6361/201220481}

\bibitem[{{Piro} \& {Bildsten}(2007)}]{Piro07}
{Piro}, A.~L., \& {Bildsten}, L. 2007, \apj, 663, 1252, \dodoi{10.1086/518687}

\bibitem[{{Remillard} {et~al.}(2021){Remillard}, {Loewenstein}, {Steiner},
  {Prigozhin}, {LaMarr}, {Enoto}, {Gendreau}, {Arzoumanian}, {Markwardt},
  {Basak}, {Stevens}, {Ray}, {Altamirano}, \& {Buisson}}]{Remillard21}
{Remillard}, R.~A., {Loewenstein}, M., {Steiner}, J.~F., {et~al.} 2021, arXiv
  e-prints, arXiv:2105.09901.
\newblock \doarXiv{2105.09901}

\bibitem[{{Revnivtsev} {et~al.}(2001){Revnivtsev}, {Churazov}, {Gilfanov}, \&
  {Sunyaev}}]{Revnivtsev01}
{Revnivtsev}, M., {Churazov}, E., {Gilfanov}, M., \& {Sunyaev}, R. 2001, \aap,
  372, 138, \dodoi{10.1051/0004-6361:20010434}

\bibitem[{{Rutledge} {et~al.}(2001){Rutledge}, {Bildsten}, {Brown}, {Pavlov},
  \& {Zavlin}}]{Rutledge01b}
{Rutledge}, R.~E., {Bildsten}, L., {Brown}, E.~F., {Pavlov}, G.~G., \&
  {Zavlin}, V.~E. 2001, \apj, 559, 1054, \dodoi{10.1086/322361}

\bibitem[{{Serino} {et~al.}(2016){Serino}, {Iwakiri}, {Tamagawa}, {Sakamoto},
  {Nakahira}, {Matsuoka}, {Yamaoka}, \& {Negoro}}]{Serino16}
{Serino}, M., {Iwakiri}, W., {Tamagawa}, T., {et~al.} 2016, \pasj, 68, 95,
  \dodoi{10.1093/pasj/psw086}

\bibitem[{{Strohmayer} \& {Bildsten}(2006)}]{Strohmayer06}
{Strohmayer}, T., \& {Bildsten}, L. 2006, in Compact stellar X-ray sources,
  Cambridge Astrophysics Series, No. 39, ed. W.~{Lewin} \& M.~{van der Klis}
  (Cambridge: Cambridge University Press), 113--156

\bibitem[{{Strohmayer} \& {Brown}(2002)}]{Strohmayer02}
{Strohmayer}, T.~E., \& {Brown}, E.~F. 2002, \apj, 566, 1045,
  \dodoi{10.1086/338337}

\bibitem[{{Sugizaki} {et~al.}(2011){Sugizaki}, {Mihara}, {Serino}, {Yamamoto},
  {Matsuoka}, {Kohama}, {Tomida}, {Ueno}, {Kawai}, {Morii}, {Sugimori},
  {Nakahira}, {Yamaoka}, {Yoshida}, {Nakajima}, {Negoro}, {Eguchi}, {Isobe},
  {Ueda}, \& {Tsunemi}}]{Sugizaki11}
{Sugizaki}, M., {Mihara}, T., {Serino}, M., {et~al.} 2011, \pasj, 63, S635,
  \dodoi{10.1093/pasj/63.sp3.S635}

\bibitem[{{VanderPlas}(2018)}]{VanderPlas18}
{VanderPlas}, J.~T. 2018, \apjs, 236, 16, \dodoi{10.3847/1538-4365/aab766}

\bibitem[{{Zhang} {et~al.}(2020){Zhang}, {Li}, {Lu}, {Song}, {Xu}, {Liu},
  {Chen}, {Cao}, {Bu}, {Chang}, {Chen}, {Chen}, {Chen}, {Chen}, {Chen}, {Cui},
  {Cui}, {Deng}, {Dong}, {Du}, {Fu}, {Gao}, {Gao}, {Gao}, {Ge}, {Gu}, {Guan},
  {Gungor}, {Guo}, {Han}, {Hu}, {Huang}, {Huo}, {Jia}, {Jiang}, {Jiang}, {Jin},
  {Jin}, {Li}, {Li}, {Li}, {Li}, {Li}, {Li}, {Li}, {Li}, {Li}, {Li}, {Li},
  {Liang}, {Liao}, {Liu}, {Liu}, {Liu}, {Liu}, {Liu}, {Liu}, {Lu}, {Lu}, {Luo},
  {Ma}, {Meng}, {Nang}, {Nie}, {Ou}, {Qu}, {Sai}, {Shang}, {Shen}, {Sun},
  {Tan}, {Tao}, {Tuo}, {Wang}, {Wang}, {Wang}, {Wang}, {Wang}, {Wang}, {Wang},
  {Wen}, {Wu}, {Wu}, {Wu}, {Xiao}, {Xiong}, {Yan}, {Yang}, {Yang}, {Yang},
  {Yi}, {Yuan}, {Zhang}, {Zhang}, {Zhang}, {Zhang}, {Zhang}, {Zhang}, {Zhang},
  {Zhang}, {Zhang}, {Zhang}, {Zhang}, {Zhang}, {Zhang}, {Zhang}, {Zhang},
  {Zhang}, {Zhang}, {Zhang}, {Zhang}, {Zhang}, {Zhao}, {Zhao}, {Zheng}, {Zhou},
  {Zhu}, {Zhu}, {Zhuang}, \& {Insight-HXMT Team}}]{hxmt}
{Zhang}, S.-N., {Li}, T., {Lu}, F., {et~al.} 2020, Science China Physics,
  Mechanics, and Astronomy, 63, 249502, \dodoi{10.1007/s11433-019-1432-6}

\end{thebibliography}
\bibliographystyle{aasjournal}

%% This command is needed to show the entire author+affiliation list when
%% the collaboration and author truncation commands are used.  It has to
%% go at the end of the manuscript.
%\allauthors

%% Include this line if you are using the \added, \replaced, \deleted
%% commands to see a summary list of all changes at the end of the article.
%\listofchanges

\end{document}